\documentclass[runningheads]{llncs}
\usepackage{graphicx}
\usepackage{cite} 
\usepackage{times}
\usepackage{epsfig}
\usepackage{amsmath}
\usepackage{amssymb}
\usepackage{mwe}
\usepackage{acro}
\usepackage{amssymb}
\usepackage{xcolor,colortbl}
\usepackage{tabularx}
\usepackage{relsize}
\usepackage{pifont}
\usepackage{booktabs} 
\usepackage{multirow}
\usepackage{multicol}
\usepackage{adjustbox}
\usepackage{float}
\usepackage[colorlinks,
            linkcolor=red,  
            anchorcolor=blue, 
            citecolor=green,   
            ]{hyperref}
\usepackage[flushleft]{threeparttable}

\begin{document}
\title{HATs: Hierarchical Adaptive Taxonomy Segmentation for Panoramic Pathology Image Analysis}
%
%

\authorrunning{R. Deng et al.}
\author{Ruining Deng\inst{1} \and
Quan Liu \inst{1} \and
Can Cui\inst{1} \and
Tianyuan Yao\inst{1} \and
Juming Xiong \inst{1}\and
Shunxing Bao \inst{1}\and 
Hao Li \inst{1} \and
Mengmeng Yin \inst{2} \and
Yu Wang \inst{2} \and
Shilin Zhao \inst{2} \and
Yucheng Tang \inst{3} \and
Haichun Yang \inst{2} \and
Yuankai Huo \inst{1,2}}


\institute{
1. Vanderbilt University, Nashville TN 37215, USA, \\
2. Vanderbilt University Medical Center, Nashville TN 37232, USA, \\
3. NVIDIA Corporation, Santa Clara and Bethesda, USA\\
}


\maketitle              
\begin{abstract}

 Panoramic image segmentation in computational pathology presents a remarkable challenge due to the morphologically complex and variably scaled anatomy. For instance, the intricate organization in kidney pathology spans multiple layers, from regions like the cortex and medulla to functional units such as glomeruli, tubules, and vessels, down to various cell types. In this paper, we propose a novel Hierarchical Adaptive Taxonomy Segmentation (HATs) method, which is designed to thoroughly segment panoramic views of kidney structures by leveraging detailed anatomical insights. Our approach entails (1) the innovative HATs technique which translates spatial relationships among 15 distinct object classes into a versatile ``plug-and-play" loss function that spans across regions, functional units, and cells, (2) the incorporation of anatomical hierarchies and scale considerations into a unified simple matrix representation for all panoramic entities, (3) the adoption of the latest AI foundation model (EfficientSAM) as a feature extraction tool to boost the model's adaptability, yet eliminating the need for manual prompt generation in conventional segment anything model (SAM). Experimental findings demonstrate that the HATs method offers an efficient and effective strategy for integrating clinical insights and imaging precedents into a unified segmentation model across more than 15 categories. The official implementation is publicly available at \url{https://github.com/hrlblab/HATs}.

\end{abstract}

\section{Introduction}

\begin{figure*}[t]
\begin{center}
\includegraphics[width=1.0\textwidth]{{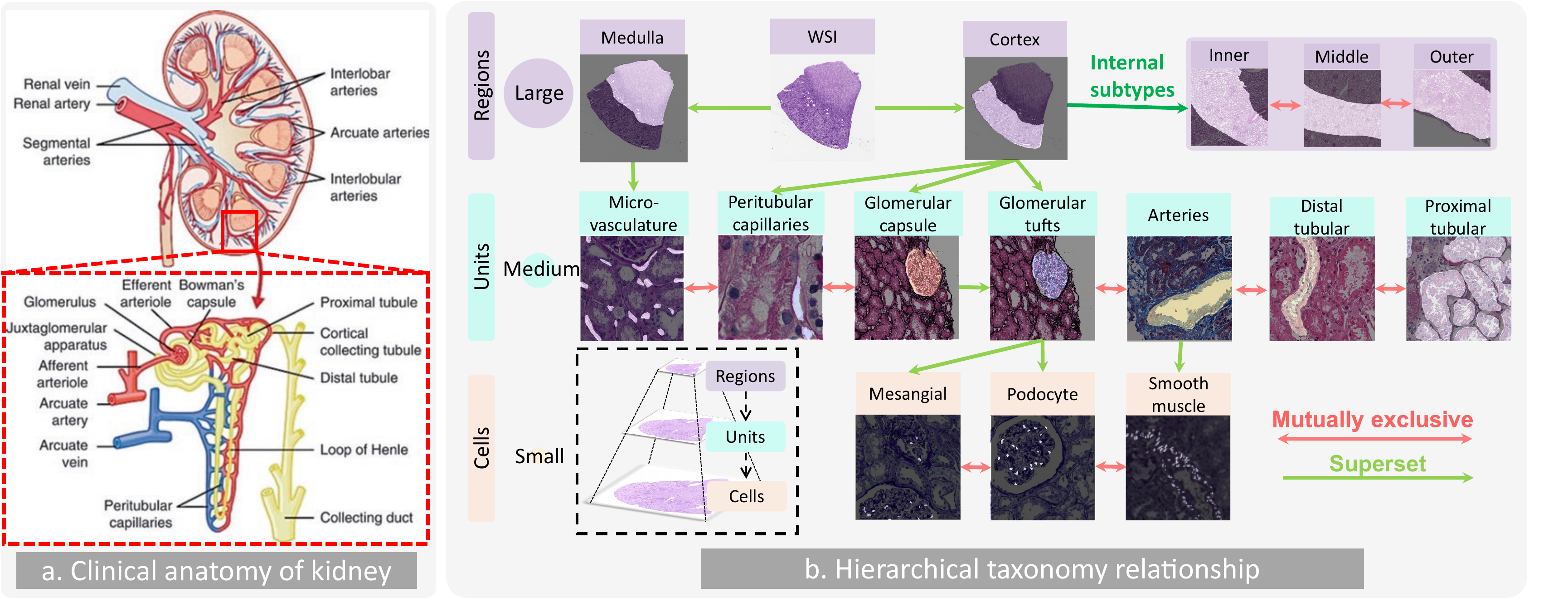}}
\end{center}
\caption{\textbf{Knowledge transformation from kidney anatomy to a hierarchical taxonomy tree.} This figure demonstrates the transformation of intricate clinical anatomical relationships within the kidney into a hierarchical taxonomy tree. (a) Pathologists examine histopathology in accordance with kidney anatomy. (b) This study revisits kidney anatomy using a hierarchical semantic taxonomy for panoramic segmentation , covering 15 classes across regions, units, and cells. The tree incorporates spatial relationships into a semi-supervised learning paradigm and uses hierarchical scale information as prior knowledge to weigh the relationship between classes.} 
\label{fig:Scope}
\end{figure*}

In renal pathology, accurate diagnosis~\cite{mounier2002cortical}, severity assessment~\cite{kellum2008acute}, and treatment efficacy~\cite{jimenez2006mast} rely on detailed examination across multiple structural levels, from broad regions (like medulla and cortex) to specific functional units (glomerulus, tubules, vessels, etc.) and individual cells. The detailed quantification across multiple organs has led to the widespread exploration of pathomics~\cite{gupta2019emergence,chen2020pathomic,chen2024spatial,huo2021ai,barisoni2020digital} as a fully quantitative approach, enhancing the current semi-quantitative clinical guidance and enabling the development of fully quantitative biomarkers. While numerous studies have advanced the segmentation of pathological images for detailed tissue analysis using deep learning techniques~\cite{kumar2017dataset, ding2020multi, ren2017computer, bel2018structure, zeng2020identification}, they face primary challenges: current architectures, which often incorporate multiple networks or heads~\cite{jayapandian2021development, li2019u, hermsen2019deep, zhang2021dodnet, wu2022tgnet, deng2023omni}, typically target individual tissue types or those within similar size ranges. These approaches lack a holistic strategy for segmentation across various anatomical levels, from broad regions to specific cells. The intricate spatial dynamics among these entities, depicted in Fig.~\ref{fig:Scope}, are crucial for comprehensive segmentation success. However, this holistic view has not been fully integrated into current deep learning advancements, leaving the complete segmentation of kidney anatomy~\cite{AL-Mamari2023} unattained.

Recently, the Segment Anything Model (SAM)~\cite{kirillov2023segment} has been proposed to provide comprehensive segmentation for everything. Many studies have endeavored to incorporate this foundational model into digital pathology~\cite{ma2024segment,deng2023segment,cui2023all}. However, there lacks a fine-tuning paradigm specifically aimed at resolving semantic segmentation challenges without explicit pixel-level prompts within foundational model architectures. 

In this work, we propose a novel Hierarchical Adaptive Taxonomy Segmentation (HATs) method, which is designed to thoroughly segment panoramic views of kidney structures by leveraging detailed anatomical insights. A hierarchical adaptive taxonomy matrix and a hierarchical scale matrix are established to translate anatomical relationships into computational modeling concepts. Moreover, The proposed method leverages the state-of-the-art AI foundation models~\cite{kirillov2023segment,ma2024segment,deng2023segment,cui2023all} and a token-based EfficientSAM~\cite{xiong2023efficientsam}. It integrates class and scale knowledge into a dynamic token bank, employing weak token prompts instead of pixel-wise ones for efficient segmentation. The contribution of this paper is threefold: 

$\bullet$ The HATs method is proposed for mathematically modeling clinical anatomy with a hierarchical taxonomy matrix and a hierarchical scale matrix for panoramic pathology segmentation. It models hierarchical spatial relationships of 15 object classes, across regions, functional units, and cells;

$\bullet$ A token-based dynamic EfficientSAM~\cite{xiong2023efficientsam}  network architecture that leverages weak token prompts to replace pixel-wise prompts to achieve superior semantic segmentation of images with partial labels, while storing class-aware knowledge and scale-aware knowledge with a token bank. 

$\bullet$ The holistic design of hierarchical matrix representation, token bank, and AI foundation model allows a single dynamic model to achieve comprehensive pathology image analysis.

\section{Methods}

The panoramic pathology segmentation comprises three integral components: (1) a hierarchical taxonomy matrix with a taxonomy loss (Fig.~\ref{fig.HATs}), (2) a hierarchical scale matrix to weight the strength of relationships in hierarchical taxonomy loss (Fig.~\ref{fig.HATs}), and (3) a dynamic EfficientSAM network with a token bank. 
\subsection{Hierarchical taxonomy matrix with taxonomy loss}

\begin{figure*}[t]
\centering 
\includegraphics[width=1.0\linewidth]{{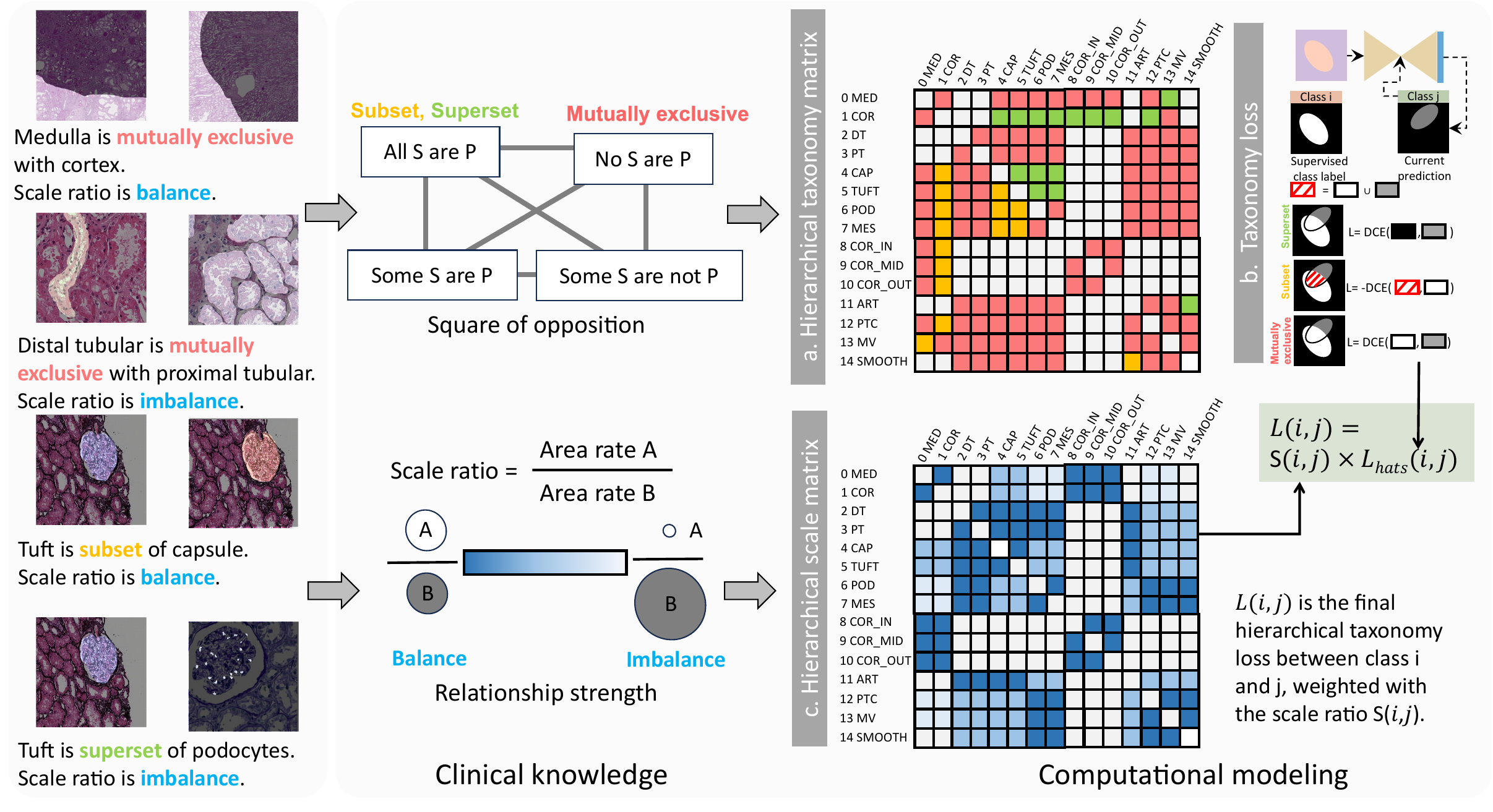}}
\caption{
\textbf{hierarchical taxonomy learning -- }
This figure highlights the key innovation of the proposed taxonomy learning strategy. (a) A hierarchical taxonomy matrix is modeled from anatomical relationships to Aristotle's logic theory in pathological image segmentation. (b) A novel taxonomy loss function is designed to operationalize the affirmative and negatory relationships from hierarchical taxonomy matrix during the training process. (c) We further encode a hierarchical scale matrix to illustrate the strength of the relationship between different objects in kidney anatomy.}
\label{fig.HATs} 
\end{figure*}

This anatomical relationship is characterized by the hierarchical taxonomy tree, as shown in Fig.~\ref{fig:Scope}b, through: \textbf{Uniqueness.} Each pair of objects is connected by a single proposition on the hierarchical tree. The expanding structure of the map, devoid of cycles, ensures stable inheritance relationships from regions down to cellular levels. \textbf{Transmissibility.} Indirect relationships between objects can be inferred from direct relationships, as established by the two fundamental categorical propositions. Relationships between objects not directly connected can be determined by combining propositions along their connecting paths on the tree. Inspired by previous work~\cite{deng2024prpseg}, we introduce an expanded 15-class hierarchical taxonomy matrix (as depicted in Fig.~\ref{fig.HATs}a), $M_t \in \mathbb{R}^{n \times n}$, to facilitate implementation in computational models. Here, $n$ represents the number of classes within the map. The matrix values includes subset ($\subseteq$), superset ($\supseteq$), and mutually exclusive ($\cap = \emptyset$).

With the introduction of the hierarchical taxonomy matrix, we incorporate spatial correlation into the training process for comprehensive segmentation using a novel taxonomy loss (as depicted in Fig.~\ref{fig.HATs}b). For a given image $I$ with a labeled class $i$, represented as $Y_i$, we generate predictions $Y'_j$ for another class $j$ within the same image. We then use the anatomical relationship to supervise the correlation between the supervised label $Y_i$ and the semi-supervised prediction $Y'_j$: (1) If $i$ is a superset of $j$, then $Y'_j$ should not exceed the region of $Y_i$; conversely, (2) if $i$ is a subset of $j$, then $Y'_j$ should cover $Y_i$ as comprehensively as possible; and (3) if $i$ and $j$ are mutually exclusive, the overlap between $Y_i$ and $Y'_j$ should be minimized. The total taxonomy loss is defined by the following equations:

\begin{equation}
    L_{\text{hats}}(i,j) = 
    \begin{cases}
    \text{DCE}(1 - Y_i, Y'_j), \quad \text{if} \quad \quad i  \subseteq j \\
    - \text{DCE}(Y_i, Y_i \cup Y'_j ), \quad \text{if} \quad i \supseteq j \\
    \text{DCE}(Y_i, Y'_j), \quad \text{if} \quad i \cap j = \emptyset \\
    0, \quad \text{if} \quad  \text{otherwise} \\
    \end{cases} 
\label{eq:taxonomyloss}
\end{equation}

\noindent where $\text{DCE}$ denotes the Dice Loss. 

\subsection{Hierarchical scale matrix with area ratio knowledge}
We formed pairwise spatial relationships across all 15 objects in the hierarchical taxonomy matrix. However, in the hierarchical taxonomy tree, different objects have varying levels of supervisory power over anatomical knowledge, depending on the size of the object and the scale of the images. For example, it is challenging to recognize explicit cells in region images at 5$\times$ magnification, while global regional knowledge is less useful and informative in cell images at 20$\times$ magnification. Therefore, we further incorporate hierarchical scale knowledge into taxonomy learning, translating hierarchical taxonomy matrix from a binary relationship~\cite{deng2024prpseg} to a fully quantitative relationship.

The hierarchical scale matrix for the 15 objects is calculated for the entire dataset to represent the strength of the relationship between two objects by their area rates, shown in Table~\ref{tab:dataset}. The area rate ($a$) for each object is determined by multiplying the pixel mean of each object in the images by the square of the micron value, and then dividing by the size of the patches. This process provides a standardized measure of an object's size in both digital and real-world dimensions.

\begin{table*}
\caption{Data collection and scale rate}
\begin{adjustbox}{width=0.95\textwidth}
\begin{tabular}{l|ccccccc}
\toprule
Class & Stain & Patch \# & Size (pixel$^{2}$) & Scale ($\times$) & Micron ($\mu$m/pixel) & Pixel mean (pixel$^{2}$) & Area rate (($\mu$m/pixel)$^{2}$)\\
\midrule
Medulla & P & 1,619 &  1024$^{2}$ & 5 & 2 & 637,975 & 2.434 \\
Cortex  & P & 3,055 & 1024$^{2}$ & 5 & 2 & 681,392 & 2.600\\
Inn. Cor.  & P & 1,242  & 1024$^{2}$ & 5 & 2 & 461,277 & 1.760\\
Mid. Cor.  & P & 1,357 & 1024$^{2}$ & 5 & 2 & 485,849 & 1.853\\
Out. Cor.  & P & 1,586 & 1024$^{2}$ & 5 & 2 & 483,486 & 1.844\\
\midrule
DT & H,P,S,T & 4,615 & 256$^{2}$ & 10 & 1 & 6,381 & 0.097   \\
PT & H,P,S,T & 4,588 &  256$^{2}$ & 10 & 1 & 23,605 & 0.360 \\
Cap.& H,P,S,T & 4,559  &  256$^{2}$ & 5 & 2 & 10,140 & 0.619  \\
Tuft& H,P,S,T &  4,536 &  256$^{2}$ & 5 & 2 & 7,641 & 0.466\\
Art.& H,P,T &  4,875 &  256$^{2}$ & 10 & 1 & 5,446 & 0.083\\
PTC& P & 4,827  &  256$^{2}$ & 40 & 0.25 & 2,152 & 0.002\\
MV& P & 1,362  &  512$^{2}$ & 20 & 0.5 & 12,905 & 0.012\\
\midrule
Pod.& P & 1,147  & 512$^{2}$ & 20 & 0.5 & 1,170 & 0.001 \\
Mes.& P & 789  & 512$^{2}$ & 20 & 0.5 & 1,079 & 0.001\\
Smooth.& P & 1,326  & 512$^{2}$ & 20 & 0.5 & 2,527 & 0.002\\
\bottomrule
\end{tabular}
\label{tab:dataset} 
\end{adjustbox}
\text{*Inn. is inner; Mid. is middle; out. is Outer; Cor. is cortex;}\\
\text{*DT is distal tubular; PT is proximal tubular; *Cap. is glomerular capsule; Tuft is glomerular tuft; }\\
\text{*Art. is arteries; MV is micro-vasculature; PTC is peritublar capillaries;}\\
\text{*Pod. is podocyte cell; Mes. is mesangial cell; Smooth. is smooth muscle}\\
\text{*H is H$\&$E; P is PAS; S is SIL; T is TRI.}
\end{table*}

With the area rate ($a$) listed in Table~\ref{tab:dataset}, we evaluate the strength of the spatial relationship in each pair of objects as the value in the hierarchical scale matrix ($S$) using the formula in Fig.~\ref{fig.HATs}c:

The total loss function is an aggregate of supervised and semi-supervised losses in~\ref{eq:loss_function}, weighted by $\lambda_{\text{hats}}$

\begin{equation}
\begin{aligned}
    L(i) = & \text{DCE}(Y_i,Y'_i) + \text{BCE}(Y_i,Y'_i)\\
    & + \lambda_{\text{hats}} \sum_{j=1}^{n}S_(i,j) \times L_{\text{hats}}(i,j) \quad ( j \neq i )
\end{aligned}
\label{eq:loss_function}
\end{equation}

\noindent where $\text{BCE}$ represents the Binary Cross-Entropy loss. $Y'_i$ is the prediction for class $i$.

\begin{figure*}[t]
\centering 
\includegraphics[width=1.0\linewidth]{{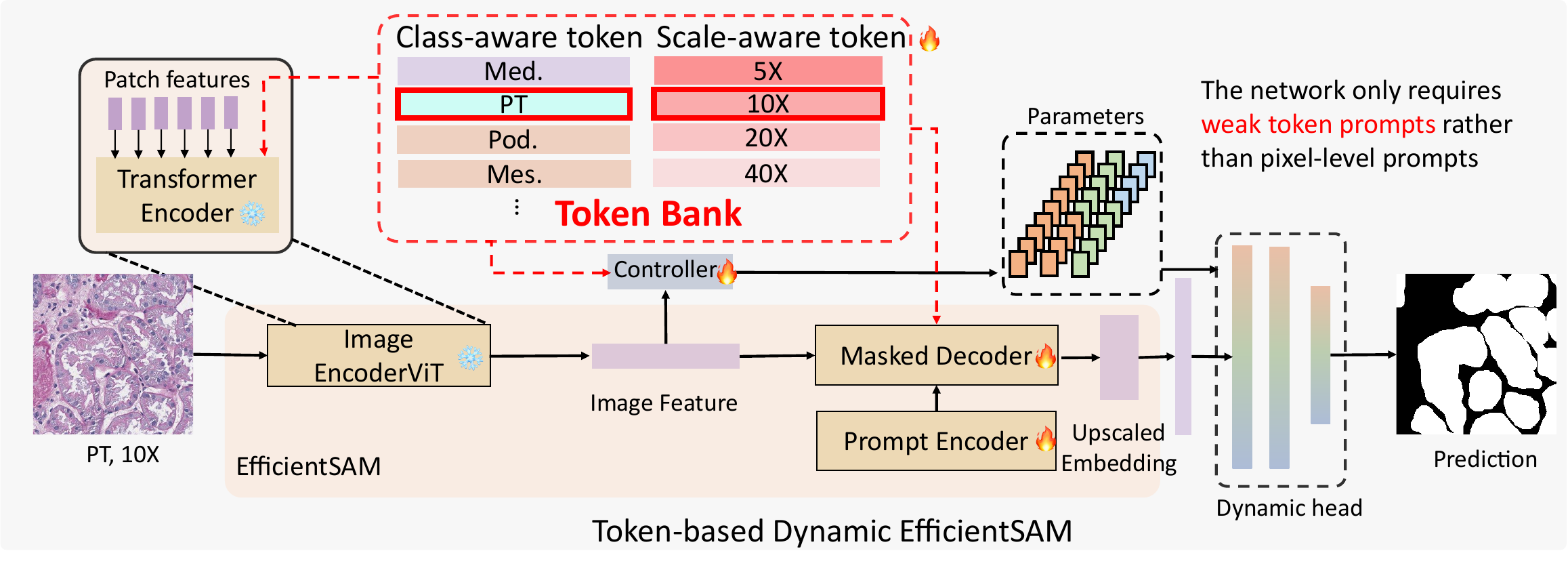}}
\caption{
\textbf{Dynamic EfficientSAM with token bank -- }
This figure visualizes the architecture of our proposed token-based dynamic EfficientSAM. Key components include a dynamic token bank with class-aware and scale-aware tokens, a token-guided imageViT encoder, a mask decoder, and a dynamic head network. This architecture leverages AI fundation model by fine-tuning with weak tokens, liberating the model from the need for pixel-level image prompts.}
\label{fig.backbone} 
\end{figure*}

\subsection{Dynamic EfficientSAM with token bank}

The architecture of dynamic EfficientSAM is presented in Fig.~\ref{fig.backbone}. The backbone of our proposed network, EfficientSAM~\cite{xiong2023efficientsam}, is chosen for its superior segmentation performance and efficient computation. Instead of using pixel-level prompts for each object in the image, a pre-defined learnable token bank is initialized to store the class-specific and scale-specific knowledge among the whole dataset. Dimensionally stable class-aware tokens ($T_c\in\mathbb{R}^{n \times d}$) and scale-aware tokens ($T_s\in\mathbb{R}^{4 \times d}$) are employed from the token bank to capture the contextual information in the model. Each class has a one-dimensional token, $t_c\in\mathbb{R}^{1 \times d}$, to store class-specific knowledge at the feature level across the entire dataset, while each magnification scale has a one-dimensional token, $t_s\in\mathbb{R}^{1 \times d}$, to provide scale-specific knowledge across four scales (5$\times$, 10$\times$, 20$\times$, and 40$\times$).

There are three module parts that use these conditional tokens to achieve semantic segmentation in the network: (1) Inspired by the Vision Transformer (ViT)~\cite{dosovitskiy2020image}, for an image $I$ of class $i$ with magnification $m$, the corresponding class token $T_c(i)$ and scale token $T_s(m)$ are stacked with the patch-wise image tokens before being fed into the current transformer block ($E_b$) (as shown in~\ref{eq:encodertoken}); (2) $T_c(i)$ and $T_s(m)$ are concatenated with latent image features ($F$) from the imageViT encoder ($E$) and a Global Average Pooling ($GAP$) in $\mathbb{R}^{d}$ for parameters ($\omega$) in the dynamic head (as shown in~\ref{eq:fusion}); (3) $T_c(i)$ and $T_s(m)$ also serve as sparse embeddings combined with dense embedding ($E_d$) in the Mask Decoder ($M_d$) to produce upscaled embeddings ($e_{upscale}$) (as shown in~\ref{eq:maskdeconder}).

\begin{equation}
\begin{aligned}
    e_b = E_b(T_c[i] || T_s[m] || e_{b-1})
\end{aligned}
\label{eq:encodertoken}
\end{equation}
\vspace{-0.5cm}
\begin{equation}
    \omega = \varphi(\text{GAP}(F) || T_c[i]  || T_s[m];\Theta_\varphi)
\label{eq:fusion}
\end{equation}
\vspace{-0.5cm}
\begin{equation}
    e_{upscale} = M_d(\text{GAP}(F) || T_c[i] || T_s[m] || E_d)
\label{eq:maskdeconder}
\end{equation}

Where $||$ represents the stacking operation, $\Theta_\varphi$ denotes the number of parameters in the dynamic head. The final semantic segmentation logits are obtained as the output of the dynamic head, inspired by~\cite{deng2023omni}.

\begin{figure*}[t]
\centering 
\includegraphics[width=1.0\linewidth]{{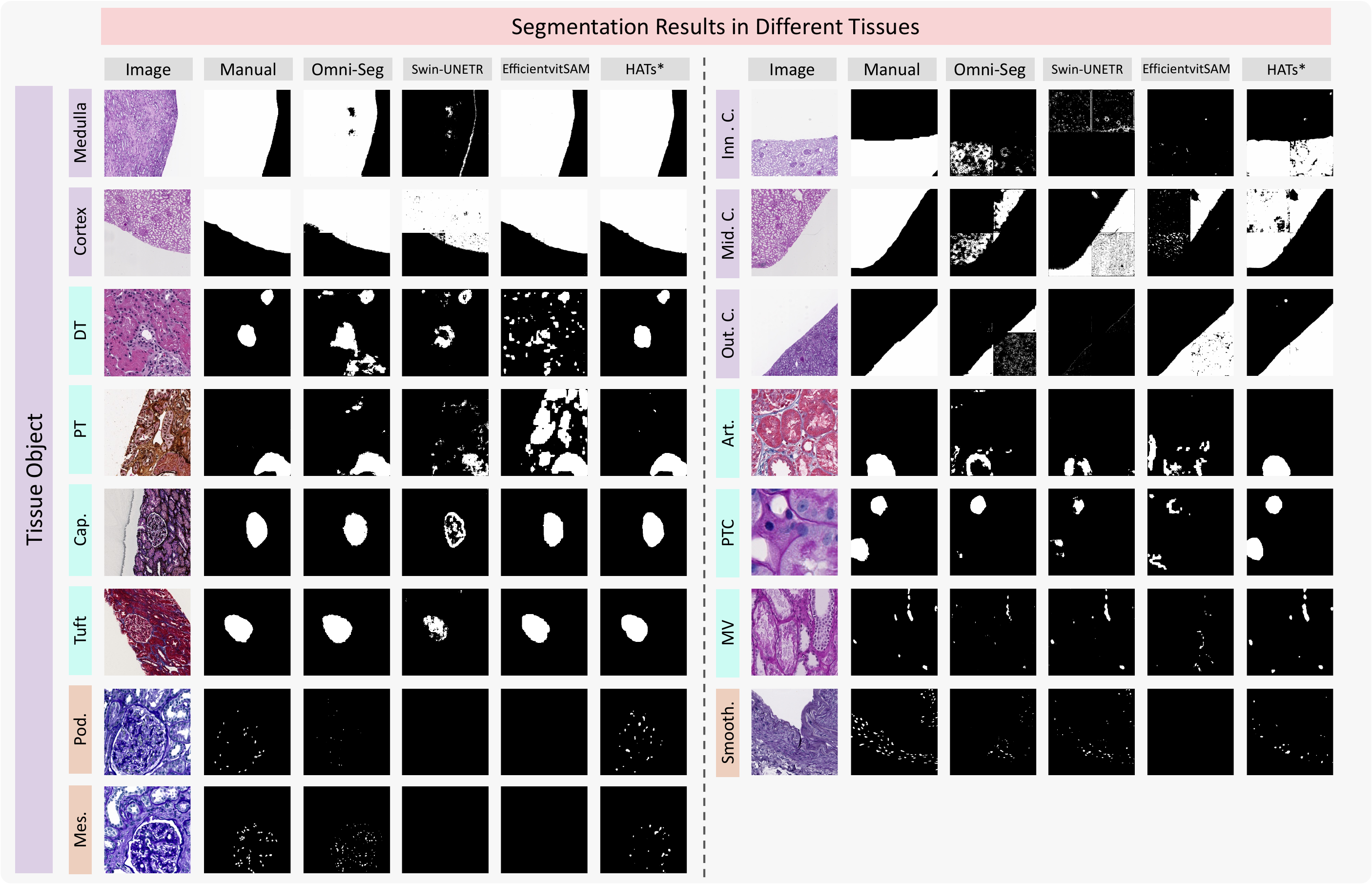}}
\caption{
\textbf{Validation qualitative results -- } This figure shows the qualitative results of different approaches. The proposed method achieved superior panoramic kidney pathology segmentation on 15 classes range regions to cells with fewer false positives, false negatives, and morphological errors.} 
\label{fig.results} 
\end{figure*}

\section{Data and Experiments}
\textbf{Data.} Our model leverages a 15-class, partially labeled dataset spanning various biological scales, from regions to cells. The dataset's structure is detailed in Table~\ref{tab:dataset}. We sourced the human kidney dataset from three distinct resources across regions, functional units and cells. Detailed data introduction can be found in the supplementary material. The dataset was partitioned into training, validation, and testing sets at a 6:1:3 ratio across all classes, with splits conducted at the patient level to prevent data leakage. 

\noindent\textbf{Experiment Details.}
The training process of our model was divided into two distinct phases. In the initial phase, which spanned the first 50 epochs, we employed a supervised learning strategy focused on minimizing binary Dice loss and cross-entropy loss. Subsequently, for the remaining epochs, both supervised and semi-supervised learning strategies were utilized, incorporating anatomy loss to explore the spatial correlation among multiple objects. All images were either randomly cropped or padded to a uniform size of $512 \times 512$ pixels prior to being fed into the model in the training stage. Testing images were initially processed using either center-cropping or non-overlapping tiling to attain the same uniform size $512 \times 512$ pixels. In our experiments, $\lambda_{\text{hats}}$ was set to 0.1. All experiments were conducted on a uniform platform, specifically a workstation equipped with an NVIDIA RTX A6000 GPU.

\section{Results}
We conducted a comparative analysis of our proposed hierarchical taxonomy learning with the dynamic EfficientSAM approach against various baseline models. These models include multi-class segmentation architectures such as (1) U-Nets~\cite{gonzalez2018multi}, (2) DeepLabV3~\cite{lutnick2019integrated}, (3) Residual-U-Net~\cite{salvi2021automated}, (4) a CNN-based multi-class kidney pathology model~\cite{bouteldja2021deep}, (5) Omni-Seg~\cite{deng2023omni}, (6) a CNN-based panoramic segmentation PrPSeg~\cite{deng2024prpseg}, (7) SegFormer~\cite{xie2021segformer}, (8) UNETR~\cite{hatamizadeh2022unetr}, (9) Swin-UNETR~\cite{hatamizadeh2021swin}, and (10) Efficient-ViT-SAM~\cite{zhang2024efficientvit}.

Table~\ref{tab:results} and Fig.~\ref{fig.results} demonstrates that our proposed method, HATs, surpasses baseline models in most evaluated metrics. Fig.~\ref{fig.results} further highlights the qualitative superiority of our approach, evidenced by reduced instances of false positives, false negatives, and morphological errors. The Dice similarity coefficient (Dice: \%, the higher, the better) was employed as the primary metric for quantitative performance assessment. The results indicate that, while multi-head designs struggle with managing spatial relationships between objects (e.g., subset/superset relationships between the capsule and tuft), the dynamic-head paradigm exhibits superior performance compared to other methods. The proposed method achieves better average performance across 15 categories.

\begin{table*}[bth]
\caption{
Performance on panoramic segmentation for kidney pathology. Dice similarity coefficient scores (\%) are reported. The difference between the reference (Ref.) method and benchmarks is statistically evaluated by Wilcoxon signed-rank test. All abbreviations are defined in Table~\textcolor{red}{1}}
\begin{center}
\begin{adjustbox}{width=1.0\textwidth}
\begin{tabular}{lc|ccccccccccccccccccc}
\toprule
\multirow{1}{0.8in}{Method} & \multirow{1}{0.8in}{Backbone} & \multicolumn{5}{c}{Regions} & \multicolumn{7}{c}{Functional units} & \multicolumn{3}{c}{Cells} & \multirow{1}{0.5in}{Average} & \multirow{1}{0.4in}{Statistic.}\\
\cmidrule(lr){3-7}
\cmidrule(lr){8-14}
\cmidrule(lr){15-17}

& & Med. & Cor. & Inn. C. & Mid. C. & Out. C. & DT & PT  & Cap. & Tufts & Art. &  PTC & MV & Pod. & Mes. & Smooth. & & \\
\midrule
U-Nets~\cite{gonzalez2018multi} & CNN & 23.87 & 64.03 & 34.53 & 32.36 & 33.62 & 47.61 & 60.45  & 45.36 & 46.62 & 47.32 & 49.21 & 48.66 & 49.92 & 49.87 & 49.77 & 45.55 & $p<$0.001\\
DeepLabV3~\cite{lutnick2019integrated}& CNN & 27.31 & 62.10 &  34.53 & 33.13 & 33.67 & 53.88 & 62.19 & 76.88 & 74.54 & 58.32 & 62.52 & 48.93 & 49.92 & 49.87 & 49.77 & 51.84 & $p<$0.001 \\
Residual-U-Net~\cite{salvi2021automated} & CNN & 24.29 & 62.24 & 34.53 & 30.53  & 47.77 & 65.76 & 78.12 & 69.62 & 79.64 & 54.74 & 60.72 & 57.29 & 66.20 & 49.87 & 56.77 & 55.87 & $p<$0.001 \\
Multi-kidney~\cite{bouteldja2021deep}& CNN & 22.81 & 68.37  & 34.54 & 30.35 & 33.60 & 67.90 & 67.70 & 85.21 & 54.87 & 55.04 & 55.66 & 68.35 &  \textbf{66.65} & 49.87  & 64.28 & 55.02 & $p<$0.001 \\
Omni-Seg~\cite{deng2023omni} & CNN & 64.91 & 70.56 & 40.16 & 36.32 & 58.07 & 63.73 & 77.44 & 87.87 & 88.00 & 56.11 & 65.22 & 55.76 & 60.30 & 62.64 & 60.93 & 63.20 & $p<$0.001\\
PrPSeg~\cite{deng2024prpseg}  & CNN & 65.21 & 70.16 & 39.52 & 36.92& 69.72 & 66.61 & 78.70 & 89.85 & 89.97 & 60.79 & 65.90 & 64.79 & 64.57 & \textbf{62.98} & 63.44 & 65.94 & $p<$0.001 \\
\midrule
SegFormer~\cite{xie2021segformer} & Transformer & 21.90 &  65.84 & 34.53 & 31.51 & 34.01 &  58.56 & 72.01 &  66.87 & 55.57 & 52.50 & 62.76 & 48.69 & 58.47 & 54.44  & 52.34 & 51.33 & $p<$0.001 \\
UNETR~\cite{hatamizadeh2022unetr} & Transformer & 23.72 & 69.43 & 34.52 & 29.57 & 33.71 & 57.54 & 71.67 & 72.14 & 51.62 & 51.36 & 54.74 & 55.86 & 49.92 & 49.87 & 49.96 & 50.37 & $p<$0.001 \\
Swin-UNETR~\cite{hatamizadeh2021swin} & Transformer & 24.13 & 69.75 & 34.33 & 29.97 & 33.83 & \textbf{68.05} & 75.40 & 78.51 & 72.59 & \textbf{65.91} & 63.74 & 68.54 & 49.95 & 49.87 & \textbf{66.54} & 56.74 & $p<$0.001 \\
\midrule
Efficientvit-SAM~\cite{zhang2024efficientvit} & SAM & 63.17 & 71.29 & 38.77 & \textbf{48.69} & 70.63 & 57.20 & 68.96 & 89.73 & 91.04 & 49.48 & 59.13 & 48.87 & 49.92 & 49.87 & 49.80  & 60.44 & $p<$0.001 \\
HATs (Ours) & SAM & \textbf{67.69} & \textbf{72.83} & \textbf{49.66} & 47.86 & \textbf{71.61} & 64.03 & \textbf{79.30} & \textbf{91.97} & \textbf{93.02} & 62.19 & \textbf{68.30} & \textbf{68.55} & 59.53 & 60.01 & 57.06  & \textbf{67.58} & \textbf{Ref.}\\
\bottomrule
\end{tabular}
\end{adjustbox}
\end{center}
\label{tab:results}
\end{table*}

\begin{table}
\caption{Ablation study of different design. Dice similarity coefficient scores (\%) are reported. The difference between the reference (Ref.) method and benchmarks is statistically evaluated by Wilcoxon signed-rank test. *HTM is Hierarchical Taxonomy Matrix with taxonomy loss, HSM is Hierarchical Scale Matrix}
\begin{center}
\begin{adjustbox}{width=0.7\textwidth}
\begin{tabular}{lccccccc}
\hline
Backbone & HTM & HSM & Regions & Units & Cells & Average & Statistic.\\
\hline
Omni-Seg~\cite{deng2023omni} & & & 53.99 & 70.59 & 61.29 & 63.20 & $p<$0.001 \\
Swin-UNETR~\cite{hatamizadeh2021swin} & & & 38.41 & 70.39 & 55.45 & 56.74 & $p<$0.001 \\
Efficientvit-SAM~\cite{zhang2024efficientvit} & & & 58.51 & 66.35 & 49.86 & 60.44 & $p<$0.001 \\
\hline
PrPSeg (CNN)~\cite{deng2024prpseg} & & &  54.96  & 73.02 & 62.77 & 64.95 & $p<$0.001 \\
PrPSeg (CNN)~\cite{deng2024prpseg} & \checkmark & &  56.31  & 73.80 & \textbf{63.64} & 65.94 & $p<$0.001 \\
PrPSeg (CNN)~\cite{deng2024prpseg} & \checkmark & \checkmark &  56.35  & 75.03 & 63.48 & 66.65 & $p<$0.001 \\
\hline
HATs (d-EfficientSAM) (Ours) &  & & 57.60   & 74.77 & 57.71 & 65.63 & $p<$0.001 \\
HATs (d-EfficientSAM) (Ours) & \checkmark &  & 60.45   & 74.98 & 58.74 & 66.89 & $p<$0.001 \\
HATs (d-EfficientSAM) (Ours) & \checkmark & \checkmark & \textbf{61.93}   & \textbf{75.34} & 58.87 & \textbf{67.58} & \textbf{Ref.} \\

\hline
\end{tabular}
\end{adjustbox}
\end{center}
\label{tab:differentdesigns} 
\end{table}

\noindent\textbf{Ablation study.} Table~\ref{tab:differentdesigns} showcases the enhancements brought about by our proposed token-based EfficientSAM and learning strategies. The results indicate that the token-based dynamic EfficientSAM generally achieves better performance in segmenting objects at all levels. With the integration of the hierarchical taxonomy matrix and hierarchical scale matrix, performance across all considered metrics is enhanced. Additionally, the performance of the proposed matrices is also evaluated with the CNN backbone from PrPSeg. The results demonstrate the generalizability and comprehensive enhancement of the matrices. However, there is a limitation: transformer-based methods exhibit better segmentation for large-scale objects (regions, objects, etc.), while CNN-based methods achieve superior performance on smaller objects (cells). It is promising to extend the current backbone by combining CNN and transformer architectures to enhance segmentation capabilities in the future work.

\section{Conclusion}
In this work, we introduce the Hierarchical Adaptive Taxonomy Segmentation method, an innovative approach for panoramic kidney structure segmentation that harnesses in-depth anatomical understanding. By formulating both a hierarchical adaptive taxonomy matrix and a hierarchical scale matrix, we successfully convert anatomical relationships into computational models. Utilizing advanced AI foundation models along with a token-based EfficientSAM, our method incorporates class and scale knowledge into a dynamic token bank, favoring weak token prompts over traditional pixel-wise prompts in the SAM-based model for enhanced efficiency in segmentation. The contributions of this study pave the way for comprehensive pathology image analysis with a single dynamic model.

\noindent\textbf{Acknowledgements}. 
This research was supported by NIH R01DK135597(Huo), DoD HT9425-23-1-0003(HCY), NIH NIDDK DK56942(ABF). This work was also supported by Vanderbilt Seed Success Grant, Vanderbilt Discovery Grant, and VISE Seed Grant. This project was supported by The Leona M. and Harry B. Helmsley Charitable Trust grant G-1903-03793 and G-2103-05128. This research was also supported by NIH grants R01EB033385, R01DK132338, REB017230, R01MH125931, and NSF 2040462. We extend gratitude to NVIDIA for their support by means of the NVIDIA hardware grant. 

\noindent\textbf{Disclosure of Interests}. The authors have no competing interests.
%
%
\bibliographystyle{splncs04}
\bibliography{main}
%




\pagebreak
%

\newcommand{\app}{\raise.17ex\hbox{$\scriptstyle\sim$}}



%
\title{HATs: Hierarchical Adaptive Taxonomy Segmentation for Panoramic Pathology Image Analysis}
\author{Supplementary Materials}
\institute{}
\maketitle              
\section{Data Introduction}
Our model leverages a 15-class, partially labeled dataset spanning various biological scales, from regions to cells. We sourced the human kidney dataset from three distinct resources:

\subsection{Regions}
Whole slide images of wedge kidney sections stained with periodic acid-Schiff (PAS, n=138) were obtained from non-cancerous regions of nephrectomy samples. The samples were categorized into several groups based on clinical data, including normal adults (n=27), patients  with hypertension (HTN, n=31), patients with diabetes (DM, n=4), patients with both hypertension and diabetes (n=14), normal aging individuals (age$>$65y, n=10), individuals with aging and hypertension (n=36), and individuals with aging, hypertension, and diabetes (n=16). These tissues were scanned at 20$\times$ magnification and manually annotated in QuPath~\cite{bankhead2017qupath}, delineating medulla, inner cortex, middle cortex, and outer cortex contours. The WSIs were downsampled to 5$\times$ magnification and segmented into 1024$\times$1024 pixel patches. Corresponding binary masks were derived from the contours.

\subsection{Functional Units}
\noindent\textbf{NEPTUNE} The distal tubular, proximal tubular, glomerular capsule, glomerular tufts, arteries, and peritubular capillaries are from the NEPTUNE study~\cite{barisoni2013digital} with 459 WSIs, encompassing 125 patients with minimal change disease, we extracted 1,751 Regions of Interest (ROIs). These ROIs were manually segmented to identify four kinds of morphology objects with normal structure and methodology outlined in~\cite{jayapandian2021development}. Each image, at a resolution of 3000$\times$3000 pixels (40$\times$ magnification, 0.25 $\mu$m per pixel), represented one of four tissue types stained with Hematoxylin and Eosin Stain(H\&E), PAS, Silver Stain (SIL), and Trichrome Stain (TRI). We treated these four staining methods as color augmentations and resized the images to 256$\times$256 pixels, maintaining the original data splits from~\cite{jayapandian2021development}.

\noindent\textbf{HuBMAP} Complementing the NEPTUNE dataset, we also incorporated data from HuBMAP. This dataset is comprised of 5 PAS-stained WSIs from varied donors, chosen based on criteria such as image quality (minimal artifacts or blurring), demographic diversity (considering age, sex, BMI), and encompassing different kidney regions (cortical, medullary, papillary). Expert segmentation was performed on the WSIs using QuPath by a lead anatomist, assisted by four other trained anatomists. They identified three types of microvascular structures: arterial/arteriole, peritubular capillary/vasa recta, vein/venule. These were later grouped under a single category termed ``microvasculature"~\cite{hubmap-hacking-the-human-vasculature}. The WSIs were then transformed into patches of dimensions 512$\times$512 at a 20$\times$ magnification.

\subsection{Cells}
We employed 17 WSIs of normal adult cases from the aforementioned nephrectomy dataset. These pathology images were scanned at 20$\times$ magnification and cropped into 512$\times$512 pixel segments to facilitate cell labeling, following the annotation process described in~\cite{deng2023democratizing}.

\end{document}